\documentstyle[prabib,aps,psfig,12pt]{revtex}
\begin{document}

\title{Experimental test for extra dimensions in Kaluza-Klein gravity}
\author{V. Dzhunushaliev
\thanks{E-Mail Address : dzhun@freenet.bishkek.su}}
\address{Universit\"at Potsdam, Institute f\"ur Mathematik,
14469, Potsdam, Germany \\
and Theor. Phys. Dept. KSNU, 720024, Bishkek, Kyrgyzstan}
\author{D. Singleton
\thanks{E-Mail Address : das3y@maxwell.phys.csufresno.edu}}
\address{Dept. of Phys. CSU Fresno, 2345 East San Ramon Ave.
M/S 37 Fresno, CA 93740-8031, USA}

\maketitle

\begin{abstract}
5D Kaluza-Klein gravity has several nonasymptotically flat
solutions which generally, possessed both electric and magnetic
charges. In this paper we suggest that these solutions
can act as quantum virtual handles (wormholes) in spacetime
foam models. By applying a sufficently large, external
electric and/or magnetic field it may be possible to ``inflate''
these solutions from a quantum to a classical state.
This effect would lead to a possible experimental
signal for higher dimensions in multidimensional gravity.
\end{abstract}

\section{Introduction}
Most modern theories which attempt to unify gravity
with the Standard Model gauge theory have extra dimensions.
These extra dimensions makes it possible to geometrize the
gauge fields (gauge bosons) according to the following
theorem \cite{per}:
\par
Let $G$ be the group
fibre of the principal  bundle.  Then  there  is a one-to-one
correspondence between the $G$-invariant metrics
\begin{equation}
ds^2 = h^2(x^\mu) (\sigma ^a + A^a_\mu dx^\mu)^2 +
g_{\mu\nu} dx ^\mu dx^\nu
\label{0}
\end{equation}
on the  total  space ${\cal X}$
and the triples $(g_{\mu \nu }, A^{a}_{\mu }, h)$.
Here $g_{\mu \nu }$ is Einstein's pseudo  -
Riemannian metric on the base; $A^{a}_{\mu }$ are the gauge fields
of the group $G$ ( the nondiagonal components of
the multidimensional metric); $h\gamma _{ab}$  is the
symmetric metric on the fibre.
\par
The off-diagonal components of multidimensional (MD)
metric act as Yang-Mills fields. One distinction between
such MD theories and 4D theories with Yang-Mills fields is that the MD
theories have a scalar field connected with the extra dimension(s).
This scalar field describes the linear size of the extra dimensions
or equivalently the volume of gauge group.
Thus one possible experimental test
for any MD gravity theory is the observation of effects arising
from this scalar field. For example, it is possible
to show that in 5D Kaluza - Klein theory
the presence of variations of the 5$^{th}$ coordinate leads to changes
in the ratio of the electrical charge to the mass of an elementary
particle. This effect is very small since no experiment has
found such a change.
\par
In this paper we offer a new possible experimental
signal for probing the extra dimensions of MD gravity
based on the existence of a certain type of spherically
symmetric nonasymptotically flat solutions
\cite {dzh0}.

\section{Wormhole and flux tube solutions in 5D gravity.}

Before discussing the 5D solutions we will briefly recall
a couple of spherically symmetric 4D electrogravity solutions
which will be used for comparison
with the 5D solutions. First there is the well known, asymptotically flat
generalized Reissner-Nordstr\"om solution which gives the gravitational and
electromagnetic fields for a point mass with both electric and magnetic
charges (the time-time component of the metric has the form
$g_{tt} = (1 - {m \over r} + {q^2 + Q^2 \over r^2})$ where $m$
is the mass and $q, Q$ are the electric and magnetic
charges respectively). Second, there is the nonasymptotically flat,
spherically symmetric Levi-Civita flux tube solution \cite{levi}
with the metric
\begin{eqnarray}
ds^2 &=& a^2\left (\cosh^2 \zeta dt^2 - d\zeta^2 - d\theta ^2 -
\sin ^2 \theta d\varphi ^2\right ),
\label{1a} \\
F_{01} &=& \rho ^{1/2} \cos\alpha, \;\;\; \; \; \; \;
F_{23} = \rho ^{1/2}\sin\alpha,
\label{1}
\end{eqnarray}
where $G^{1/2} a \rho ^{1/2} = 1$ ;
$\alpha$ is an arbitrary constant angle; $a$ and
$\rho$ are constants defined by Eq. (\ref{1a}) - (\ref{1});
$G$ is Newton's constant ($c=1$, is the speed of light);
$F_{\mu\nu}$ is the electromagnetic field tensor.
Both the generalized Reissner-Nordstr\"om solution
and the Levi-Civita flux tube solution place no restrictions
on the relative values of the electric and magnetic
charges.
\par
In 5D Kaluza - Klein theory there are
intriguing wormhole (WH) and flux tube solutions \cite {dzh0}
\cite{dzh1}. Here we give a brief summary of these solutions.
The general form of the metric is:
\begin{eqnarray}
ds^2 &=& e^{2\nu (r)}dt^{2} - r_0^2e^{2\psi (r) - 2\nu (r)}
\left [d\chi +  \omega (r)dt + n\cos \theta d\varphi \right ]^2
\nonumber \\
&-& dr^{2} - a(r)(d\theta ^{2} +
\sin ^{2}\theta  d\varphi ^2),
\label{3}
\end{eqnarray}
where $\chi $ is the 5$^{th}$ coordinate;
$\omega = A_t$ and $n \cos \theta = A_{\phi}$
are the 4D electromagnetic potentials; $n$ is an integer;
$r, \theta, \varphi$ are ``polar'' coordinates. The
5D spacetime is the total space
of the U(1) principal bundle, where the fibre is the
U(1) gauge group and the base is ordinary 4D spacetime.
\par
A detailed analytical and numerical investigation
of the metric in Eq. (\ref{3}) gives the following
spacetime configurations, whose global structure
\textit{depends} on the relationship
between the electric and magnetic fields \cite{dzh1}:
\begin{enumerate}
\item
$0 \le H_{KK} < E_{KK}$. The corresponding solution
is \textbf{\textit{a WH-like object}}
located between two surfaces at $\pm r_0$ where the
reduction from 5D to 4D spacetime breaks
down\cite{chodos}. The cross-sectional
size of this solution (given by $a(r)$) increases as $r$ goes from $0$
to $\pm r_0$. The throat between the $\pm r_0$ surfaces is
filled with electric and/or magnetic flux.
As the strength of the magnetic field increases the longitudinal
distance between the surfaces at $\pm r_0$  increases.
This can be seen diagrammatically from the first
two pictures in Fig.\ref{fig1}.
\begin{figure}
\centerline{
\framebox{
\psfig{figure=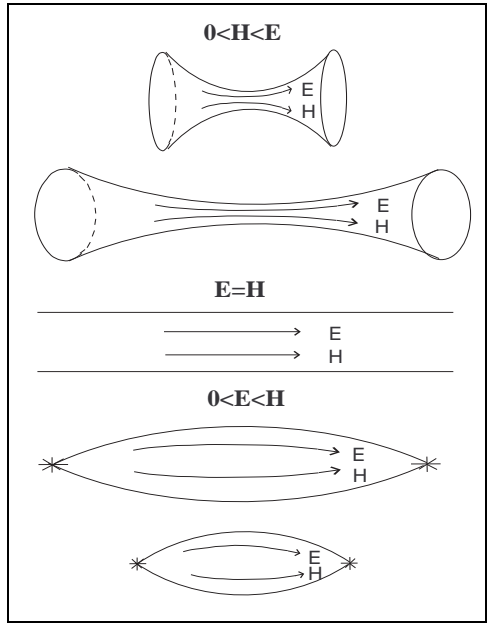,height=8cm,width=6cm}}}
\vspace{5mm}
\caption{The evolution from WH-like solution
to finite flux tube solution}
\label{fig1}
\end{figure}
\item
$H_{KK} = E_{KK}$. In this case the solution is
\textbf{\textit{an infinite flux tube}} filled
with constant electrical and magnetic fields,
with the charges disposed at $\pm \infty$. The cross-sectional
size of this solution is constant ($ a= const.$). In Refs.
\cite{dzh1} \cite{dzh3} an exact, analytical form of this
solution was given in terms of hyperbolic functions. This
solution if almost identical to the 4D Levi-Civita flux
tube solution except the strength of the magnetic and
electric fields are equal, while in the Levi-Civita
solution the two fields can take on any relative value
with respect to one another. The restriction that the
electric charge equals the magnetic charge is reminiscent
of other higher dimensional soliton solutions. In
Ref. \cite{perry} non-Abelian, Kaluza-Klein dyon solutions
were found which obeyed the same restriction that
the ``electric'' charge equal the ``magnetic'' charge.
The present flux tube solution can be viewed as two
connected or bound Kaluza-Klein dyons.
The form of this infinite flux tube
configuration also has similarities to
the Anti-de Sitter (AdS) ``throat region''  that one
finds by stacking a large number of D3-branes \cite{jm}.
Both the spacetime around the D3-branes and the
electric/magnetic flux tube have indefinitely long
cylindrical ``throats'' which can be thought of as ending
either on the horizon of a black hole (for the D3-branes solution),
or on an electric/magnetic charged object (for the flux tube solution).
\item
$0 < E_{KK} \le H_{KK}$. In this case we have
\textbf{\textit{a finite flux tube}}
between two (+) and (-) magnetic and/or electric
charges, which are located at $\pm r_0$. The longitudinal
size of this flux tube is finite, but now the cross
sectional size decreases as $r \rightarrow r_0$. At
$r = \pm r_0$ this solution has real singularities which
we interpret as the locations of the magnetic and/or
and electric charges. The behavior of this flux tube
solution as $E_{KK}$ decreases can be seen diagrammatically
from the last two pictures in Fig. 1
\end{enumerate}

\section{The basic idea}

\begin{figure}
\centerline{
\framebox{
\psfig{figure=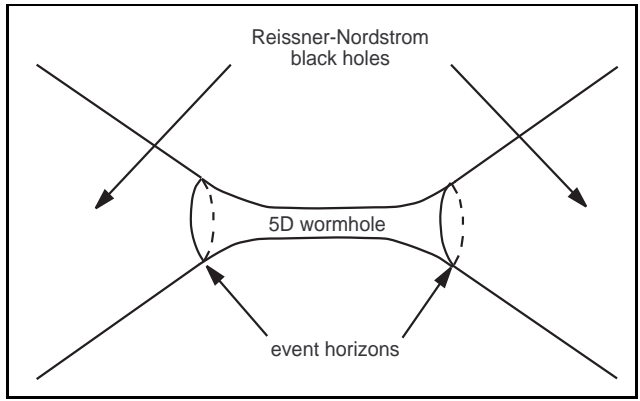,height=5cm,width=8cm}}}
\vspace{5.mm}
\caption{The forces acting on 2 ends of WH}
\label{fig2}
\end{figure}

In Ref.\cite{pirt} the idea was advanced that
\textit{\textbf{a piecewise compactification mechanism}} can
exist in Nature. Piecewise compactification implies
that some parts of the Universe
are regions where one has full MD gravity (5D in our case),
while other parts of the Universe are
ordinary 4D regions where gravity does not act
on the extra dimensions. For this mechanism to be viable it
is necessary that on the boundary between these regions
a quantum splitting off of the 5$^{th}$ dimension occurs.
In regions where gravity propagates in all the dimensions
the Universe will appear as a true 5D spacetime.
\footnote{in this case the fifteen 5D Einstein vacuum
equations = 4D gravity + Maxwell electrodynamic +
scalar field.}
In the regions where gravity does not propagate
into the extra dimension one has ordinary
4D spacetime plus the gauge fields of the fibre.
\footnote{in this case there are fourteen 5D Einstein vacuum equations =
4D gravity + Maxwell electrodynamic, where $G_{55}$ = scalar field
does not vary.}. The boundary between these
regions should be Lorentz invariant surfaces
\footnote{for the 4D observer at infinity it will
appear as an event horizon.}.
An example of such a construction is the composite WH of
Ref. \cite{dzh3} which consists of two 4D Reissner - Nordstr\"om
black holes attached to either end of a 5D WH solution
(see Fig.\ref{fig2}).
\par
The proposed experimental signal for the extra dimensions in
MD gravity relies on these postulated composite WH structures.
The basic idea is the following:
\textit{\textbf{Composite WHs can act as a quantum handles
(quantum WHs) in the spacetime foam. These quantum
structures can be ``blown up'' or ``inflated'' from a
quantum state to a classical state by embedding it in parallel
$E$ and $H$ fields with $E > H$.}}
These quantum handles are taken as quantum fluctuations
in the spacetime  foam which the externally imposed
$E$ and $H$ fields can then promote to classical states with
some probability.
This process is envisioned as taking place inside a solenoid
which has an additional electric field, $E$, parallel to
magnetic field $H$.
\par
The above process has some similarity to the pair
production of electric or magnetic charged black holes in an
external electric or magnetic field \cite{ross} \cite{hawking}.
In Ref. \cite{hawking} it was shown that despite the fact that the
Maxwell action ($F_{\mu \nu} F^{\mu \nu} = {\bf H} ^2 - {\bf E}^2$)
changes sign under a dual transformation of ${\bf H}$ and
${\bf E}$ that the pair production of electric black holes
and magnetic black holes are identical {\it and} suppressed.
In the next section we consider an approximate flux tube solution
which has ${\bf E} \approx {\bf H}$ and therefore has a Maxwell
action which is approximately zero.

\section{A more detailed description}

\begin{figure}
\centerline{
\psfig{figure=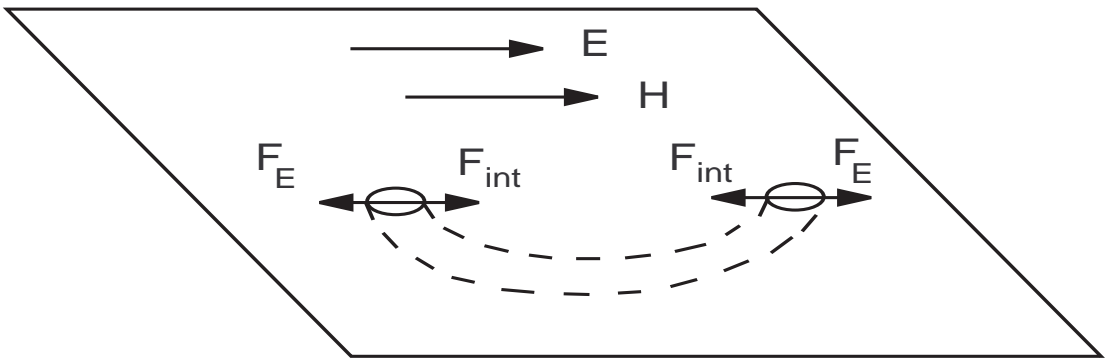,height=5cm,width=15cm}}
\vspace{5.mm}
\caption{The forces acting on 2 ends of WH}
\label{fig3}
\end{figure}

The first three pictures in Fig. 1 represent
solutions where the charges are unconfined and separated
by some finite, longitudinal distance. For an external
observer these composite WHs will appear as two oppositely
charged electric/magnetic objects, with the charges
located on the surfaces where the 4D and 5D
spacetimes are matched. Since one would like these
electric/magnetic charged objects to be well separated,
we will consider the case $E \approx H$ $(E>H)$.
(This leads to the Maxwell action, $F^2 = H^2 - E^2$,
being approximately zero). Under these condition
the solution to Einstein's MD vacuum equations
for the metric ansatz given in Eq. (\ref{3}) is \cite{dzh1}
\cite{dzh4}
\begin{eqnarray}
q \approx Q,
\label{4}\\
a \approx \frac{q^2}{2} = const,
\label{5}\\
e^{\psi} \approx e^{\nu} \approx \cosh \left( \frac{r\sqrt{2}}{q} \right),
\label{6}\\
\omega \approx \frac{\sqrt{2}}{r_0}\sinh \left( \frac{r\sqrt{2}}{q} \right)
\label{7}
\end{eqnarray}
here $q$ is the electrical charge and $Q$ the magnetic charge.
Both Kaluza - Klein fields are:
\begin{equation}
E \approx H \approx \frac{q}{a} \approx \frac{2}{q} 
\approx \sqrt{\frac{2}{a}}
\label{8}
\end{equation}
The cross sectional size of the WH is proportional to
$q^2$. According our scenario the external, parallel electric
and magnetic fields should fill the virtual WH. Changing
Eqs. (\ref{4})-(\ref{7}) into \textit{cgs}
units the electric and magnetic fields necessary for
forming a composite WH with a cross sectional size
$a$ is
\begin{equation}
E \approx H \approx \frac{c^2}{\sqrt G} \sqrt{\frac{2}{a}}
\label{9}
\end{equation}
From Eq. (\ref{9}) it can be seen that the larger the
cross sectional size, $a$, of the WH the smaller the
$E$ and $H$ fields. However, in order for the charged surfaces
of the WH to appear as well separated electric/magnetic
charged objects we need to require that the longitudinal
distance, $l$, between these surfaces be much larger than
the cross sectional size of the WH, $l \gg \sqrt{a}$.
Also in order to be able to separate
the two ends of the WH as distinct electric/magnetic
charged objects one needs the external force to be much
larger than the interaction force between the
oppositely charged ends. This leads to the following condition
which is illustrated in Fig.\ref{fig3}. 
\begin{eqnarray}
F_{ext} &=& qE + QH \gg \frac{q^2 + Q^2}{l^2} = F_{int} \nonumber \\
&\approx& 2qE \gg \frac{2 q^2}{l^2} \rightarrow l \gg \sqrt a
\label{10}
\end{eqnarray}
If this condition holds than the oppositely charged ends will
move apart. Otherwise the ends will come back together
and annihilate back into the spacetime foam.
\par
The average value of $a$ for spacetime foam is given by the Planck size
${\sqrt {a}} \approx L_{Pl} \approx 10^{-33}$ cm.
Thus the relevant field $E$ should be
$E \approx \sqrt{2 c^7} / G \sqrt{\hbar} \approx 3.1 \times 10^{57} V/cm$.
This field strength is in the Planck region, and
is well beyond experimental capabilities to create.
Hence one must consider quantum WHs
whose linear size satisfies $\sqrt a \gg L_{Pl}$. The
larger $\sqrt a$ the smaller the field strength needed.
But such large quantum WHs are most likely very rare. If $f(a)$ is the
probability density for the distribution for a
WH of cross section $a$
then $f(a)da$ gives the probability for the appearance of
quantum WH with cross section $a$. The bigger
$a$ the smaller the probability, $f(a)da$.
Also the larger the value of $E$ and $H$ the smaller is
the cross sectional size $a$ of the WH that can be
inflated from the spacetime foam. Thus depending on the
unknown probability $f(a) da$ one can set up some spatial region
with parallel $E$ and $H$ fields whose magnitudes
are as large as technologically feasible, and look for
electric/magnetic charged objects whose charges are of
similar magnitude. Finally, it has been proposed
\cite{hamed} that the Planck scale may occur at a much lower
energy scale ($\approx 10^3 $ GeV) than is normally thought
($\approx 1 \times 10^{19}$ GeV) due to the presence of large,
extra dimensions. In such a scenario ${\sqrt {a}} \approx L_{Pl}
\approx 10^{-18}$ cm, and the field strength would decrease
by fifteen orders of magnitude so that $E \approx 3.1 \times
10 ^{42}$ from above. This is still beyond experimental
capabilities, however now one can consider quantum WHs
that are of a smaller size, $a$, as compared to the standard
case when the Planck size is $10^{-33}$ cm. Combining the large
extra dimension scenario with the inflation of
the electric/magnetic infinite flux tube
solution by external fields, tends to increase the probability
of observing such an event.
\par
The energy density $u$ of electrical and magnetic fields
stored in such an inflated WH is
\begin{equation}
u = \frac{E^2}{8\pi} + \frac{H^2}{8\pi}
\approx \frac{E^2}{4\pi}  = \frac{1}{4\pi}
\frac{c^4}{G}\frac{2}{a}.
\label{11}
\end{equation}
In this case the energy $U$ is
\begin{equation}
U \approx \pi a l w = \frac{c^4}{2G} l
\label{12}
\end{equation}
$U$ increases linearly with $l$ as one would expect for two
objects connected by a flux tube. This
places a restriction on $l$, since as $l$ increases
beyond a certain point the energy will be large
enough to favor creating another electric/magnetic
charged pair.

\section{Conclusion}

We have presented a possible experimental scheme
to test the presence of the higher dimensions
in MD gravity through the use of certain WH-like solutions,
and an assumption about piecewise
compactification on the surface where the reduction from
5D to 4D breaks down.
\par
The difference between the present solutions the
4D Levi-Civita flux tube solution is that
the 5D solution requires that the magnitudes of
the electric and magnetic charges be of the same
magnitude. This restriction on the charges is
similar to that for certain non-Abelian, Kaluza-Klein
dyon solutions \cite{perry}. In the 4D case any relative strength
between the charges is allowed. For both the 4D and 5D
solutions it can be
asked how the Dirac condition between electric/magnetic
charges fits into all of this. Recently
an investigation into closely related cosmic magnetic flux tube solutions
was carried out \cite{dav}. It was found that in the
context of these GR solutions the Dirac condition
is modified so that (magnetic flux) + (dual electric charge)
is the quantized object rather than just the
magnetic flux.

\section{Acknowledgments}
VD is supported by a Georg Forster Research Fellowship
from the Alexander von Humboldt Foundation and H.-J. Schmidt
for invitation to Potsdam Universit\"at f\"ur research.

\end{document}